# Adaptive user support in educational environments: A Bayesian Network approach

*A. G. Stoica*

University «Dunărea de Jos » of Galaţi
Faculty of Economic & Administrative Sciences
Str. Nicolae Bălcescu nr. 59-61, Galaţi 6200, România
ace@k.ro

*N.K Tselios, C. Fidas*

University of Patras
ECE Department – HCI Group
GR-26500 Rio Patras, Greece
nitse@ee.upatras.gr, fidas@clab.ee.upatras.gr

SUMMARY

This paper is concerned with the design and implementation of an innovative user support system in the frame of an open educational environment. The environment adapted is ModelsCreator (MC), an educational system supporting learning through modelling activities. The pupils' typical interaction with the system was modelled using Bayesian Belief Networks (BBN). This model has been used in ModelsCreator to build an adaptive help system providing the most useful guidelines according to the current state of interaction. A brief description of the system and an overview of application of Bayesian techniques to educational systems is presented together with discussion about the process of building of the Bayesian Network derived from actual student interaction data. A preliminary evaluation of the developed prototype indicates that the proposed approach produces systems with promising performance.

INTRODUCTION

During the last years a number of *open problem-solving environments* have been built that are based on the constructivist approach. Because of the nature of these environments, user interaction in their context could be very rich and the patterns of use of the tools that are included in them cannot be fully anticipated. The complexity of such environments can often lead to poor usability, which can be an obstacle to obtaining the expected pedagogical value from their use. Improving usability of these environments is an objective that can be achieved in various ways. For instance various usability evaluation techniques that take in consideration both the pedagogical value and the usability of the environments have been proposed [1,13]. Also more complex task modelling approaches have been proposed for the design of such environments [16]. Also the support at run time to the user through *adaptive user support systems* is an approach that can improve usability. It should be recognised that overall artificial intelligence techniques have not succeeded to deliver the expected results in the educational field, through the Intelligent Tutoring Systems. However the premise of adaptive system behaviour through which higher usability and increased system transparency can be obtained, remains a valid scientific objective. In recent years the heavy modelling involved in traditional AI approaches has been replaced by implicit models built from rich data sets through techniques proposed by the machine learning and knowledge discovery fields. These techniques have produced during the last years efficient algorithms and found new areas of applications. Among them a technique that has certain advantages and has been extensively used during the last years are the Bayesian Belief Networks (BBN's) [11]. A Bayesian Belief Network [14] is a directed acyclic graph where each node represents a random variable of interest and each edges represents direct correlations between the variables. Because of the underlying probabilistic model that describes the belief on the existence of a specific event, BBNs are considered one of the strongest ways to represent uncertainty. By capturing decisions on accurate cognitive models of the users and then modelling uncertainty in human computer interaction, the modelling of the user, the user's interface behaviour and, therefore, the efficiency of the interaction can be improved. Bayesian reasoning is based in formal probability theory and is used extensively in several current areas of research, including pattern recognition and classification. Assuming a random sampling of events, Bayesian theory supports the calculation of more complex probabilities from previously known results [10]. The advantages of BBNs include the simple process for constructing probabilistic networks even from relative a small amount of data, the efficient algorithms to evaluate probabilities for instances of a node and the versatile knowledge representation which such networks provide.

During the reported experiment we have attempted to build such an adaptive user support system for an open educational environment using Bayesian Networks. In this paper we first present an overview of Bayesian probabilistic theory together with a brief description of BBNs and their implementation in educational environments. Then a short description of ModelsCreator is included, i.e. the system, which was used as a platform for applica-tion of the developed technique. Then the proposed approach to adapt the user support module through a BBN, which was constructed from a large amount of log files of actual usage of the system is presented. Finally, a preliminary system testing and evaluation experiment that took place in order to demonstrate the benefits of the proposed architecture is presented, followed by a short presentation of the results obtained from our experience of student modelling with BBN in open problem solving environments.

BAYESIAN MODELING

The underlying mathematical theory of BBNs consists mainly of Bayes' formula in various forms and generalizations. The Fundamental Rule of probability calculus is the following:

$$P(a|b)P(b)=P(a,b),$$

where $P(a,b)$ is the probability of the joint event $a \cap b$.

From this follows $P(a|b)P(b) = P(b|a)P(a)$ and this yield the well known Bayes' Rule:

$$P(b|a) = P(a|b)P(b)/P(a)$$

Learning Bayesian Belief Networks from data

One of the hardest tasks in Bayesian modelling is finding the most probable dependency model [7]. Let us see how we get the right model having the data.

Let us denote our dependency model with the letter M and our data with D. Applying Bayes' rule to this we can calculate the probability of the dependency model given the data D as follows:

$$P(M|D)=P(D|M)P(M)/P(D) \quad (1)$$

Here $P(M)$ is the probability of the model before we have taken into consideration any observation data. In these state it is impossible to calculate $P(M|D)$ because we cannot calculate $P(D)$, the «general» probability of the data. But as we can observe the term $P(D)$ appears in the expressions that denote probability for any model M so we can calculate the ratio between the probabilities of two models $M_1$ and $M_2$ as follows:

$$P(M_1|D)/P(M_2|D)=[P(D|M_1)P(M_1)/P(D)] / [P(D|M_2)P(M_2)/P(D)] \quad (2)$$

By applying equation (2) to (1) we get :

$$P(M_1|D)/P(M_2|D)= [P(D|M_1)P(M_1)]/[P(D|M_2)P(M_2)] \quad (3)$$

Since the probabilities are never negative, ratio $P(M_1|D)/P(M_2|D)$ being greater than one can only happen if $M_1$ is more probable than $M_2$. Similarly this ratio can be less than one if $M_1$ is less probable than $M_2$.

$P(M)$ is the prior probability of the model. A common way to determine prior probabilities for data analysis purposes is to try to let the data « speak » and not to favour any models beforehand. So for this reason we assume that all models have the same prior probability meaning that for any two models $M_1$ and $M_2$ we have the following :

$$P(M_1)=P(M_2) \quad (4)$$

So now our formula is further simplified because $P(M_1)$ and $P(M_2)$ cancel out and we get :

$$P(M_1|D)/P(M_2|D)= P(D|M_1)/P(D|M_2) \quad (5)$$

So at last we have only to calculate $P(D|M)$. $P(D|M)$ can be calculated by the following formula [5]:

$$P(D|M) = \prod_{i=1}^{n}\prod_{j=1}^{q_i} \frac{\Gamma(\frac{N'}{q_i})}{\Gamma(\frac{N'}{q_i}+N_{ij})} \prod_{k=1}^{r_i} \frac{\Gamma(\frac{N'}{r_i q_i}+N_{ijk})}{\Gamma(\frac{N'}{r_i q_i})}$$

After defining the Bayesian network, inference algorithms should be defined that reason following this formula in order to update the beliefs. There are two categories of inference algorithms: *exact algorithms* and *stochastic sampling algorithms* [7]. Probably the most used exact algorithm is clustering algorithm. Even though the clustering algorithm is the fastest exact algorithm available, there are networks for which the memory requirements or the updating time obtained by using them may be not acceptable. In these cases, the researcher may decide to sacrifice some precision and choose an approximate sampling algorithm. Sampling algorithms are based on a statistical technique known as Monte Carlo simulation, in which the model is run through individual trials involving deterministic scenarios. The final result is based on the number of times that individual scenarios were selected in the simulation.

USE OF BBN IN EDUCATIONAL SYSTEMS

Various prototypes have been produced demonstrating that BBNs are suitable for effective modelling of student behaviour [8]. BBNs have been utilized in various ways to achieve adaptability in educational environments in terms of determining student goals, determining feedback, curriculum sequencing and fine-tuning the pedagogical strategy to deliver knowledge. ANDES [3], a system to teach physics problem solving techniques to college students, uses BN to identify the current problem solving approach of the user. ANDES also use a BBN to determine what hints to provide to the user by identifying how the student is solving a problem and how he has progressed down the solution path. In CARME (an open problem solving environment to teach geometrical concepts to pupils of the age group 11-14), a BBN has been used to infer problem-solving strategies that the pupil applied in order to solve a given problem from low-level interactions [15]. In this case, identification of user's strategies has been used to facilitate classification of the solutions presented by the students and aiding the evaluation of the learning process and the quality of interaction presented to the users.

In addition to reasoning at a lower interaction level about student actions it is possible to use a BBN in order to reason at a higher level. Collins et al. [2] constructed a BBN that represents a hierarchy of skills for arithmetic, containing the test questions, the theory and the user's skill in specific topics. The semantics implied by the links represent which question refers to a specific topic. The system considers the current estimate of the student's ability to compute the probability of responding correctly to each question contained in the database schema. Then the system select an item that the student has a near to 50% chance of answering correctly. Regarding the selected pedagogical approach, the criteria of item selection could be altered appropriately. CAPIT, [12] is a constraint-based tutor for English capitalization and punctuation that uses BBN for long term student modelling. An evaluation of this system showed that a group of students using the adaptive version learned the domain rules at a faster rate than the group that used the non-normative version of the same system.

As discussed in this section, probabilistic models, such as BBNs have focused in great extend on traditional "drill and test" systems. In this paper we argue that complex, non-linear interaction educational systems such as ModelsCreator could benefit as well from a BBN approach

USING A BBN IN MODELS CREATOR

The Models Creator Environment
Open problem solving environments[1] are computer-based learning environments that let students actively explore certain concepts while they are engaged in problem solving, with emphasis in the active, subjective and constructive character of learning [17]. Compared to traditional learning environments the student's activity cannot be reduced in a sequence of pre-defined tasks so the student has the freedom to explore various entities in her own unique way.
ModelsCreator (MC) is such an environment that permits modelling activities to young students [9]. Two main design principles were specified: to support expression through *different kinds of reasoning* in a simplified and synthetic mode and model mechanisms that derive from different subject matters, which permit *interdisciplinary* approaches. Models can be built in MC, using qualitative reasoning, as well as quantitative and semi-quantitative relations that can be used to link various objects, representing primitive concepts.

---
[1] Another term often used is that of "microworlds"

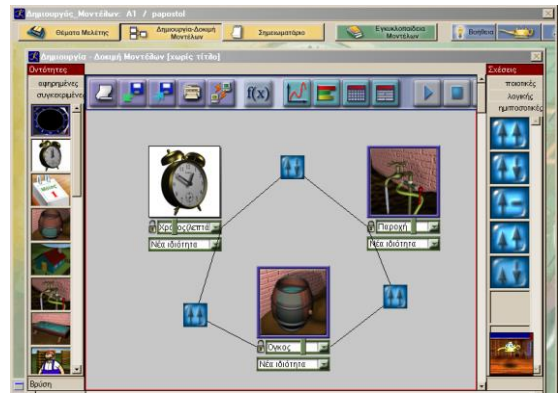

*Figure 1:* The modelling environment of Models Creator.

The system supports high level of *visualisation*, combines modelling tools with real world simulations, and incorporates alternative and multiple forms of representation [4]. Concerning the MC interface design, a direct manipulation interface suitable for young students has been designed, satisfying the criteria of minimising of the distance of execution and the distance of meaning. An extract of this interface is shown in figure 1, in which the modelling environment of MC can be seen.

It has been proven [9] that this system can provide a rich and constructive learning experience to young students.

MC consists of seven inter-related components (Figure 2): *"Study Themes"*, "*Modelling Space*", *"Notebook"*, *"Models' Encyclopaedia"* "*Communication system*", "*Files' management system*" and "*Help system*".

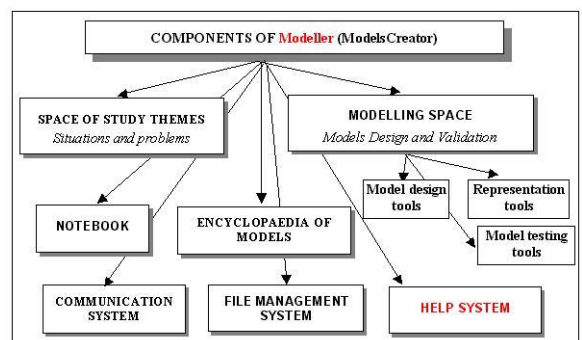

*Figure 2:* The architecture of Models Creator.

Online help adaptation using Bayesian networks
A BBN has been developed in the context of the reported research, that was used in order to adapt the user support (HELP) system of ModelsCreator. The BBN was built using data collected from real world evaluations of the MC environment, The model was based on the selected data, so the approach we followed can be considered as a data-centric one. The data used to train the structure and the probabilities of the Bayesian network was obtained from a quite large amount of log files [6] produced by ModelsCreator during previous experiments, involving 15 students of 11-14 years age group. Representative tasks were given to the students to accomplish. An extract of the log files used can be seen in figure 3.

A log file pre-processing procedure took place first that resulted in building of a database containing some 2200 logged actions. Our effort was to construct a BBN to obtain a belief about the next most possible action of the student with respect to the current state of the interaction flow. So every row in the database, after the pre-processing procedure, contained the following fields : Previous Action (Paction), Previous Property (Pprop) , Previous Action Time, Current Action (Caction), Current Property (Cproperties), Current Action Time, Next Action (Naction), Next Property and the difference between current and previous time (CPTime_d). The database has this structure since the temporal ordering of the actions has not been taken into account. The range of the possible models that can be build using MC is very large so there is almost no constraint in solutions that the user might build to a certain problem. Thus, the adaptation is very difficult to be done in a higher cognitive goal level (for example concerning specific goal which the user attempts to accomplish, or even a sub-goal as a part of a pupil's approach to solve a problem). So our approach focuses on general patterns that can occur in the stream of user interaction with the software. After we tried several configurations for the data used in order to train the Bayesian network, we discovered that best results were obtained when we kept track of previous and current action and we tried to guess the next action.

```
00 : 08 : 14  InsertObject            Plant
00 : 08 : 19  ChooseAttribute         Growth->Plant
00 : 08 : 37  InsertObject            Leaf
00 : 08 : 40  InsertObject            Sun
00 : 08 : 46  ChooseAttribute         Photosynthesis->Leaf
00 : 08 : 49  ChooseAttribute         Intensity of Light->Sun
00 : 09 : 07  InsertRelation          Proportional
00 : 09 : 07  ConnectRelation         Photosynthesis
00 : 09 : 07  ConnectRelation         Intensity of Light
00 : 09 : 48  InsertRelation          Proportional
00 : 09 : 48  ConnectRelation         Growth
00 : 09 : 48  ConnectRelation         Photosynthesis
00 : 10 : 00  RunModel
00 : 11 : 19  BarChartActivation
00 : 11 : 21  GraphChooseAttribute Growth->Plant
00 : 11 : 23  GraphChooseAttribute Intensity of Light->Sun
00 : 11 : 24  GraphChooseAttribute Photosynthesis->Leaf
00 : 11 : 31  RunModel
```

*Figure 3* : Extract of the log file used to build the BBN.

In the derived BBN (figure 4), causal relationships between previous action and current action, and current action and next action have been depicted. From an interaction flow perspective, the structure is semantically meaningful and the relation between the previous and current action of a student is an expectable empirical model of interaction To find the probability of occurrence of an instance of the next action variable, only the current instance of the current action is needed. The structure of our Bayesian network is a tree structure denoting that it is suitable to be used in an on-line training as well because this structure implies that the evaluation time for this network is polynomial.

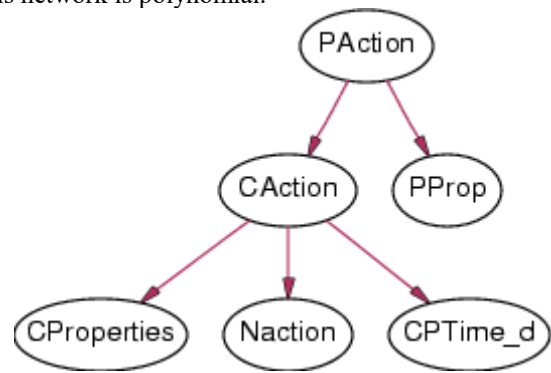

*Figure 4:* The final Bayesian network obtained.

Another advantage of the data-centric approach used to construct a Bayesian network against efficiency-centric and expert-centric is that its predictive performance can be certified by testing the network against an amount of data that was not used to train it [12]. Indeed, by using the ten fold cross validation method, the performance of the network was measured to be 88,43% (+-1,36%, p<0.05) which is considered remarkably high considering the limited amount of data used to train the network.

Environment architecture modification
The architecture of ModelsCreator was enriched with a new module that incorporated the developed BBN and enhances the functionality of the User Support Module (Help System). The new module was called Adaptive User Support Module (AUSM), This new architecture is presented conceptually in figure 5.

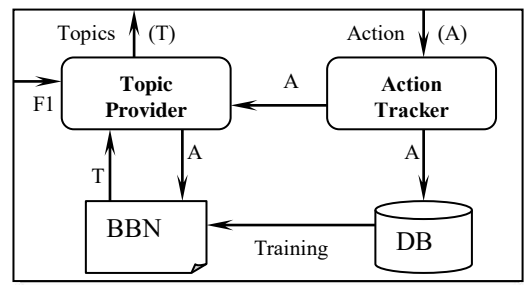

*Figure 5:* Modules interaction in the frame of the Enhanced User Support System of MC.

The role of the new module is twofold:

- to collect data from the stream of the user actions and to write them in a database.
- to provide at any time the most appropriate help topics according to the probabilistic calculus made using the developed BBN.

The database obtained is used as input for a Bayesian modelling tool that will process the data in the database and will construct the BBN that models our problem. The resulting network is exported in *Hugin Lite* file format. The module parses the file that contains the network and uses the conditional probability tables in it to provide the most useful help topics to ModelsCreator by predicting the most probable actions of the student to follow. This Bayesian network is actually used by a module that interacts with our open problem-solving environment, ModelsCreator and with a Bayesian modelling tool through files (it reads the BBN description file and it writes the database with the data to update the probabilities of the network).

At run time the developed module operates as follows: Every action of the user it sent to the new AUSM module through an appropriate method call. When the user asks for help the ModelsCreator will call another method of AUSM, that will return the most probable topics for the current state of the interaction. These actions are obtained by instantiating in the Bayesian network the current action node and calculating the probabilities for the next action instances. Each next action is related to a relevant help topic. When a new action is performed the AUSM module does the following: it writes in the database the data relating to the previous, current and next actions, then it sends further the action performed so that the module is able to instantiate the current action node with the current action performed providing this way the adaptive version of user support.

PRELIMINARY EVALUATION EXPERIMENT
The developed system evaluation was done through an experiment. As described in the previous section, the Bayesian network structure and conditional probabilities were derived from data collected from 15 log files produced by actual use of the ModelsCreator software.

A new log file was obtained during an experiment in which the task to complete was of different nature than that of the tasks in the training data set. This log file contained 141 records of user actions (including the properties and time attributes as discussed above) and was used to evaluate and test the adaptive module. An example of the actual behaviour of the AUSM module is showed in figure 6. When a user tried to connect unsuccessfully two objects with a relation and asked for help, the topic explaining how to use 'connect relation' appeared.

The module was used by a program that simulated the performing of the actions from this last log file and logged the action and the help topics provided by the AUSM module.

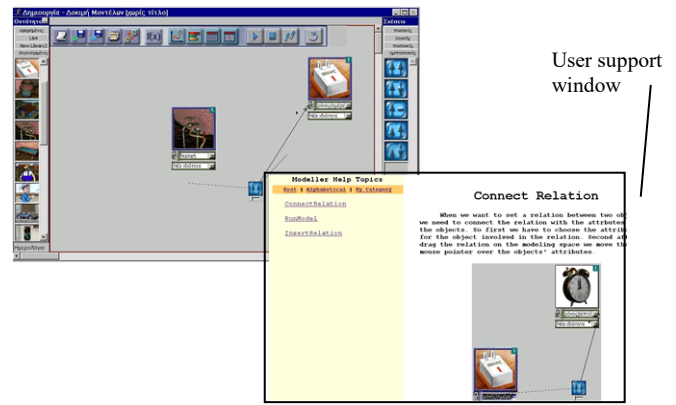

*Figure 6:* The user interface of the adaptive user support environment, 3 options are provided to the user in the help window, with the most probable one actually shown to the user

For every action performed the module provides the most probable three next actions. This approach is to make easier the selection that the user has to do – instead of selecting from a range of many topics concerning a variety of actions the user is provided with the most probable three. The help topics for every action are presented in the form: most probable next action, second most probable action and third most probable action.

The experiment results were the following :
The adaptive module guessed the next action and designated as the most probable action in 44,681% of the cases, as the second most probable in 24,113% of the cases and as the third most probable action in 8,511% of the cases. So the overall result is that the adaptive module guessed correctly the next action in 77,305% of the cases. The results are of the same order with those produced by the mathematical evaluation of the BBN derived, especially if we consider the fact that the task presented here was completely different from the tasks that actually the BBN was constructed from.

CONCLUSIONS
This paper presents the development of an adaptive user support system (AUSM) for an open learning environment. The architecture of AUSM was based on a BBN. The performance of the developed module during the preliminary evaluation experiment was very promising. The system, when fed with interaction data of a new problem solving task was able to predict correctly 2 out of 3 actions and therefore to provide the user potentially with useful support if required in an efficient way.

It is argued that a more efficient and adaptive user system increases indirectly the pedagogical value of the en-vironment because it contributes to the transparency of the tool, a more intuitive flow of interaction and enhances the learnability of complex tools such as those included in open problem solving environments. Our approach could be expanded to adapt the behaviour of the open educational environment in various aspects such as goal recognition, adaptive assessment regarding the expertise of the user, and adaptation of the interface (for example, adaptation of right click pull down menus, together with a constant set of commands). Further research Is needed in order to investigate these areas, together with more exhaustive evaluation of the effects of the developed module and its effect on the way it affects student learning within such an environment.

ACKNOWLEDGEMENTS
Financial support has been provided by the ModelsCreator/Pinelopi Program of the Greek Ministry of Education, for development of MC software. Also the Erasmus/EU program has funded the first author during the last year who has been with the HCI Group of the University of Patras. Special thanks are also due to V. Komis of Univ. of Patras and A. Dimitracopoulou of Univ. of the Aegean for permitting use of log data and the MC environment. Also special thanks are due to Dan Neagu who started and inspired the long-standing Greek-Romanian collaboration. This paper is dedicated to him.

BIBLIOGRAPHY
[1] Avouris, N., M., Tselios, N.,K , Tatakis E.,C., D*evelopment and Evaluation of a Computer-based Laboratory Teaching Tool*, Journal Computer Applications in Engineering Education, 2000 (accepted for publication).

[2] Collins J., Greer J., Huang S., *Adaptive Assessment Using Granularity Hierarchies and Bayesian Nets*. In Proceedings of Intelligent Tutoring Systems , pages 569-577,1996.

[3] Conati C., Gertner A.S. , VanLehn K., Druzdel M., *On-line Student Modeling for Coached Problem Solving Using Bayesian Networks*. In Proceedings of the Seventh International Conference on User Modeling, pp 231-242,1997.

[4] Dimitracopoulou, A., Komis, V., Apostolopoulos, P., Politis P., *Design principles of a new modelling environment for young students, supporting various types of reasoning and interdisciplinary approaches*. International Conference on Artificial Intelligence in Education, Le Mans, France,1999.

[5] Glymour C. and Cooper G. (eds.). Computation, Causation & Discovery. AAAI Press/The MIT Press, 1999

[6] Gudzial, M. J., *Deriving Software Usage Paterns From Log Files* Georgia Institute of Technology. GVU Center Techical Report.Report #93-41,1993.

[7] Heckerman, D., *A Tutorial On Learning Bayesian Networks.* Technical Report MSR-TR-95-06, Microsoft Research, November, 1996

[8] Jameson A., *Numerical uncertainty management in User and Student Modeling : An Overview of Systems and Issues*. In User Modeling and User-Adapted Interaction 5, 1995.

[9] Komis V., Dimitracopoulou A., Politis P., Avouris N., Expérimentations exploratoires sur l'utilisation d'un environnement informatique de modélisation par petits groupes d'élèves, Sciences et Techniques Educatives, Vol. 8, no 1-2, April 2001, pp.75-86.

[10] Luger, G. F., Stubblefield, W. A., *Artificial Intelligence – structures and strategies for complex problem solving,* Third Edition, Addison Wesley, 1998.

[11] Niedermayer, D., *An Introduction To Bayesian Networks And Their Contemporary Applications* , University of Saskatchewan, Technical Report 184-3-54440 1998.

[12] Mayo M., Mitrovic A, Optimizing *ITS Behaviour with Bayesian Networks and Decision Theory*. In International Journal of Artificial Intelligence in Education , 2000 (12).

[13] Squires D., & Preece, J., *Predicting Quality in Educational Software: Evaluating for Learning, Usability and the Synergy between them*, Interacting with Computers, 11, 1999 , pp. 467-483.

[14] Stephenson, T. A., *An Introduction To Bayesian Network Theory And Usage ,* Institut Dalle Molle d'Intelligence Artificielle Perceptive, Technical Report IDIAP-RR 00-03, 2000.

[15] Tselios N. K., Maragoudakis M., Avouris N. M., Fakotakis, N., Kordaki, M. (2001) *Automatic diagnosis of student problem solving strategies using Bayesian Networks* (In Greek) , 5th Panhellenic conference in mathematics and informatics in education, Thessaloniki, 12-14 October 2001.

[16] Tselios, N.K , Avouris ,N.M., Kordaki M., *Student task modeling in design and evaluation of open-problem solving environments*, Journal of Education and Information Technologies(submitted for publication), 2001.

[17] Von Glasersfeld, E. *Learning as a constru-tive activity*. In Janvier, C. (Ed.) Problems of representation in teaching and learning of mathematics London: Lawrence Erlbaum associates, 1987, pp. 3-18.